\documentclass[twocolumn,useAMS,usegraphicx,usenatbib,aas_macros]{mn2e}

\usepackage{array} 
\usepackage{hyperref}
\usepackage{amsfonts}
\usepackage{amssymb} 
\usepackage{amsmath}
\usepackage{graphicx}
\usepackage{enumerate}
\usepackage{color}
\usepackage{mathrsfs}
\usepackage{comment} 
\bibpunct{(}{)}{;}{a}{}{,} 

\title[Modelling fully convective stars]{Modelling fully convective stars in eclipsing binaries:\\
KOI-126 and CM Draconis}
\author[F. Spada and P. Demarque]
 {  F.~Spada$^1$ and P.~Demarque$^1$  \\
$^1$ Department of Astronomy, Yale University, 260 Whitney Avenue, New Haven, CT 06511, USA \\
}

\begin{document}

\def\aj{{AJ}}                   
\def\araa{{ARA\&A}}             
\def\apj{{ApJ}}                 
\def\apjl{{ApJ}}                
\def\apjs{{ApJS}}               
\def\aap{{A\&A}}                
\def\apss{{Ap\&SS}}          
\def\aapr{{A\&A~Rev.}}          
\def\aaps{{A\&AS}}              
\def\mnras{{MNRAS}}             
\def\nat{{Nature}}              
\def\ssr{{Space~Sci.~Rev.}}

\date{}

\pagerange{\pageref{firstpage}--\pageref{lastpage}} \pubyear{}

\maketitle

\label{firstpage}

\begin{abstract}
We present models of the components of the systems KOI-126 and CM Draconis, the two eclipsing binary systems known to date to contain stars with masses  low enough to have fully convective interiors.  
We are able to model satisfactorily the system KOI-126, finding consistent solutions for the radii and surface temperatures of all three components, using a solar-like value of the mixing-length parameter $\alpha$ in the convection zone, and PHOENIX NextGen 1D  model atmospheres for the surface boundary conditions. 
Depending on the chemical composition, we estimate the age of the system to be in the range $3$--$5$ Gyr.
For CM Draconis, on the other hand, we cannot reconcile our models with the observed radii and $T_{\rm eff}$ using the current metal-poor composition estimate based on kinematics. 
Higher metallicities lessen but do not remove the discrepancy.  
We then explore the effect of varying the mixing length parameter $\alpha$. 
As previously noted in the literature, a reduced $\alpha$ can be used as a simple measure of the lower convective efficiency due to rotation and induced magnetic fields.
Our models show a sensitivity to  $\alpha$ (for $\alpha < 1.0$) sufficient to partially account for the radius discrepancies.  
It is, however, impossible to reconcile the models with the observations on the basis of the effect of the reduced $\alpha$ alone.
We therefore suggest that the combined effects of high metallicity and $\alpha$ reduction could explain the observations of CM Draconis. 
For example, increasing the metallicity of the system towards super-solar values (i.e. $Z=2\,Z_\odot$) yields an agreement within $2\,\sigma$ with $\alpha  = 1.0$. 
\end{abstract}


\section{Introduction}
 
Low-mass stars (i.e., $M \lesssim 1~M_\odot$) represent the largest fraction of the stellar population in the Galaxy. 
The availability of a reliable theoretical mass--radius relationship is of considerable interest both {\it per se} and in view of the application to the study of exoplanet-hosting stars. 
Moreover, below about $0.30~M_\odot$, the depth of the subsurface convection zone extends enough to reach the centre and stars become fully convective (hereafter FC). 
This new regime has peculiar characteristics and presents additional challenges.
In constructing theoretical models of FC stars, special care is required in handling all the main input physics ingredients, such as the equation of state, the opacities, and the atmospheric boundary conditions, due to the lower temperature and higher density regime (see the reviews by \citealt{Allard_ea:1997} and \citealt{Chabrier_Baraffe:2000}).
Although convection becomes very efficient at quite shallow depths (i.e. $\nabla \approx \nabla_{\rm ad}$ at $T \gtrsim T_{\rm eff}$, see e.g. section~7.3.3 of \citealt{Kawaler}), it is still necessary to rely on the mixing length theory (MLT, \citealt{BV58}) for the convective transport in the outermost layers and in the atmosphere, which can have a significant impact on the global structure parameters \citep{Chabrier_Baraffe:1997}. 

Double-lined eclipsing binary systems (hereafter EBs) provide very accurate determination of the fundamental parameters of stars, allowing to test theoretical model predictions.
There is mounting evidence that low-mass stars in EBs have radii about $10$ per cent larger and effective temperatures about $5$ per cent cooler than theoretical models (see \citealt{Ribas:2006, Ribas_ea:2008,Morales_ea:2009} and references therein).
Interestingly, the discrepancies are in the right proportions to produce roughly the same luminosity. 
To explain these results, \citet{Chabrier_ea:2007} proposed that close EBs may not be representative of the whole low-mass stars population because of the high rotation regimes produced by spin--orbit synchronization via tidal interaction.
This explanation is corroborated by the significant correlation found by \citet{LopezMorales:2007} between the difference in the radii and the level of magnetic activity. 
Indeed, both rotation and magnetic fields can reduce the efficiency of convection, which is measured by the parameter $\alpha$ in the standard MLT framework \citep[e.g.][]{Tayler:1973}. 
A faster rotation, moreover, can enhance the magnetic activity through the dynamo effect, leading to a substantial increase in the star-spots coverage of the surface.
Both effects were phenomenologically taken into account by \citet{Chabrier_ea:2007}, using a lower $\alpha$ with respect to the solar-calibrated value and reducing the radiative flux at the surface proportionally to the star-spots filling factor $\beta$, respectively.
Their results show that the parameters $\alpha$ and $\beta$ have a degenerate effect on the stellar radius. 
They also claim that FC stars are almost insensitive to a variation of $\alpha$ and very low values (e.g., $\alpha \lesssim 0.1$) are required to reconcile the models with the observations. 
Finally, according to \citet{Morales_ea:2010}, the discrepancy may be partially accounted for by a systematic error in the light curve-derived radii, which would occur if a value of $\beta$ of about $35$ per cent and a concentration of the spots near the stellar poles are postulated.
As \citet{Burrows_ea:2011} have pointed out, however, theoretical models calculated with a quite restricted set of parameters (e.g., solar metallicity), have been compared with observations of stars of poorly determined or even unknown metallicity, assuming a general validity that is not rigorously justified. 
Even a moderate increase in metallicity can result in theoretical radii increases which are significant in comparison with the discrepancies discussed so far.
A combination of all these effects is thus the most likely solution to the low-mass stellar radii problem.
  
In this paper, we attempt to model the components of the two EBs known to date to host FC stars. 
Our main focus is to determine the sensitivity of the best-fitting models to various possible choices of the parameters, with as much generality as possible.
The paper is organized as follows: in Section~\ref{sec:obs} we discuss the available data and compare the physical properties of the two systems; in Section~\ref{sec:models} we present the code and input physics used in our calculations and in Section~\ref{sec:ttau} we show the impact of the atmospheric boundary conditions on our models.
In Sections~\ref{sec:k126} and \ref{sec:cmdra} we discuss our best-fitting solutions for the systems KOI-126 and CM Draconis (hereafter CM Dra), respectively. 
We compare our results for KOI-126 with those obtained by \citet{Feiden_ea:2011} in Section~\ref{sec:comp}. 
Finally, in Section~\ref{sec:disc}, we review results and conclusions of this work.

\section{The systems KOI-126 and CM Dra}
\label{sec:obs}

The systems KOI-126 (KIC number: $5897826$) and CM Dra host the only EBs known to date in the FC mass range.
The former was recently discovered by the Kepler mission \citep{Carter_ea:2011}; it is a hierarchical triple system, with a low-mass binary subsystem (orbital period: $1.767$ d) orbiting the primary star with a period of $33.92$ d. 
The second system (e.g. \citealt{Lacy:1977}, \citealt{Morales_ea:2009}, and references therein) is composed of two FC stars (orbital period: $1.268$ d) which have common proper motion with a white dwarf. 

For the stars in these systems, masses and radii are determined with very high precision from the modelling of the light curve eclipses. 
Other parameters however, such as metallicity and effective temperatures, are much less well known.
Table \ref{tab:parms} summarises the known physical parameters of the two systems.

Spectroscopic observations of KOI-126 only show features associated with the primary star \citep[][see also the online supporting material]{Carter_ea:2011}.
However, the light curve analysis provides a determination of the fluxes of the two low-mass stars relative to the primary, which can be used to derive a rough estimate of their effective temperatures. 
We proceeded as follows: using synthetic spectra from the NextGen (NG) grid\footnote{available online at \texttt{ftp://phoenix.hs.uni-hamburg.de/NextGen/}.}
\citep[constructed with the PHOENIX 1D code,][]{Hauschildt_ea:1999}, with solar metallicity and appropriate $\log g$ and $\log T_{\rm eff}$, we calculated the theoretical integrated flux for each star of the system, taking into account the response function of Kepler \citep{VanCleve_Caldwell:2009}.
We estimate that the effective temperatures fall in the range $3200$--$3300$ K; since the radii are known, the luminosities can be derived as well (see Table \ref{tab:parms}).

Although considerable effort has been devoted to the study of CM Dra, both spectroscopically and photometrically (\citealt{Lacy:1977}, \citealt{Metcalfe_ea:1996}), the direct determination of its metallicity is still controversial \citep{Viti_ea:1997, Viti_ea:2002, Morales_ea:2009}; the system is considered a Population II member based on its measured proper motion.

\begin{table}
\begin{center}
\caption{Physical parameters of the two systems. The KOI-126 data marked with a $^\dag$ are estimates based on the quoted flux ratio of each component to the primary: $f_B/f_A = 3.26\pm 0.24 \cdot 10^{-4}$, $f_C/f_A=2.24 \pm 0.48 \cdot 10^{-4}$ (see text). The orbital periods are: $P_{ABC}=33.92$ d (KOI 126 B-C binary around KOI-126 A); $P_{BC}=1.767$ d (KOI-126 B-C binary); $P_{12}=1.268$ d (CM Dra). The metallicity range of CM Dra reported here is that given in the literature.} 
\scalebox{0.95}{
\begin{tabular}{cc|cc}
\hline
\hline
 \multicolumn{2}{c}{KOI-126} &  \multicolumn{2}{c}{CM Dra}   \\
 \hline
 \multicolumn{4}{c}{Mass $[M_\odot]$} \\
$M_A$ & 1.347 $\pm$ 0.032  & - & - \\
$M_B$ & 0.2413 $\pm$ 0.0030 & $M_1$ & 0.2310 $\pm$ 0.0009   \\
$M_C$ & 0.2127 $\pm$ 0.0026 & $M_2$ & 0.2141 $\pm$ 0.0010  \\
\hline
 \multicolumn{4}{c}{Radius $[R_\odot]$} \\
$R_A$ & 2.0254 $\pm$ 0.0098 & - & - \\
$R_B$ & 0.2543 $\pm$ 0.0014 & $R_1$ & 0.2534 $\pm$ 0.0019 \\
$R_C$ & 0.2318 $\pm$ 0.0013 & $R_2$ & 0.2396 $\pm$ 0.0015 \\
\hline
 \multicolumn{4}{c}{Log. surface gravity, $\log_{10} (g/{\rm cm^2\,s^{-1}})$} \\
$\log g_A$ & 3.9547 $\pm$ 0.0069 & - & - \\
$\log g_B$ & 5.0101 $\pm$ 0.0029 & $\log g_1$ & 4.994 $\pm$ 0.007 \\
$\log g_C$ & 5.0358 $\pm$ 0.0027 & $\log g_2$ & 5.009 $\pm$ 0.006 \\
\hline
 \multicolumn{4}{c}{Effective temperature $[K]$} \\
$T_{\rm eff,A}$ & 5875 $\pm$ 100  & - & - \\
$T_{\rm eff,B}$ & ($3300$)$^\dag$  & $T_{\rm eff,1}$ & 3130 $\pm$ 70  \\
$T_{\rm eff,C}$ & ($3200$)$^\dag$  & $T_{\rm eff,2}$ & 3120 $\pm$ 70  \\ 
\hline
 \multicolumn{4}{c}{Luminosity} \\
$\log \frac{L_A}{L_\odot}$ & 0.6417 $\pm$ 0.029 & - & - \\
$\log \frac{L_B}{L_\odot}$ & ($ -2.14$)$^\dag$ & $\log \frac{L_1}{L_\odot}$ & $-2.258$ $\pm$ 0.038 \\
$\log \frac{L_C}{L_\odot}$ & ($ -2.26$)$^\dag$  & $\log \frac{L_2}{L_\odot}$ & $-2.313$ $\pm$ 0.056 \\
\hline
 \multicolumn{4}{c}{Metallicity} \\
$[{\rm Fe}/H]$ & $+0.15$ $\pm$ 0.08 & $[M/H]$ & $-1.0<[M/H]<-0.6$ \\
\hline
\end{tabular}
}
\label{tab:parms}
\end{center}
\end{table}

\begin{figure}
\begin{center}
\includegraphics[width=0.5\textwidth]{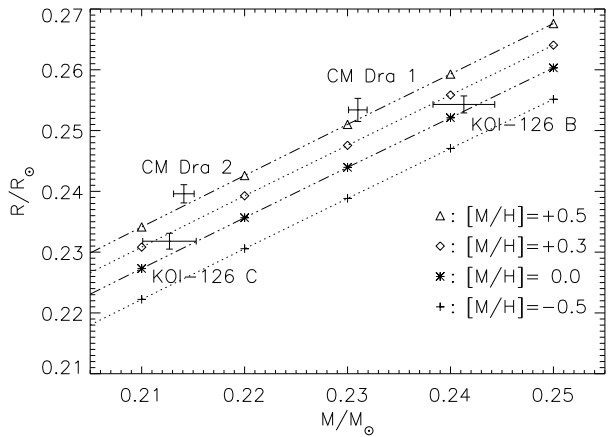}
\caption{Observed masses and radii for the FC stars (showing the error bars) in the two systems compared with theoretical mass--radius tracks at $4$ Gyr, calculated with different metallicities and $\alpha=2.0$.}
\label{fig:mrrel}
\end{center}
\end{figure}

The masses of the FC components of both systems fall in the range $0.20<M/M_\odot<0.25$ and have remarkably similar ratios.
Interestingly, KOI-126 C and CM Dra 2 have the same mass within the error, but the radius of the latter is significantly larger (of about $3$ per cent).
Similarly, KOI-126 B is about $4$ per cent more massive than CM Dra 1, but their measured radii are the same within the error.
The mass--radius relationship followed by CM Dra 1 and 2 is indeed very different from that of the FC components of KOI 126, as Fig. \ref{fig:mrrel} shows.
Correspondingly, the effective temperatures of CM Dra are significantly lower than the range estimated for KOI-126 B and C. 
From the raw observational data, taken at face value, we can expect the best-fitting models for the FC stars in the two systems to have quite different properties.

In modelling the two systems, we used different strategies to cope with their different characteristics and available observational constraints (e.g., spectroscopy of KOI-126 A, $T_{\rm eff}$ of CM Dra 1 and 2, etc.).
For KOI-126, we looked for a self-consistent solution (in age and composition) for all the three components, without making any \textit{a priori} assumption on the age, as is explained in detail in Section~\ref{sec:k126}.
For CM Dra, on the other hand, we adopted the age estimate from the literature, since the age sensitivity of FC models is too weak to provide a meaningful self-consistency check (e.g., the radius variation is of the order of $2$--$3$ per cent over the whole main sequence lifetime, $1$--$10$ Gyr, and about $1$ per cent between $1$ and $5$ Gyr).

\section{Construction of the stellar models}
\label{sec:models}
  
All the models were calculated using the Yale Rotating Stellar Evolution Code (YREC, see \citealt{Demarque_ea:2008}) in its non-rotating configuration. 
We used the \citet{Ferguson_ea:2005} opacities at low temperatures, the OPAL Rosseland opacities at high temperatures \citep{Iglesias_Rogers:1996}, and the OPAL 2005 equation of state \citep{Rogers_Nayfonov:2002}. 
The energy generation rates are calculated according to the prescription of \citet{Bahcall_Pinsonneault:1992} and the diffusion coefficients for helium and heavy elements are those given by \citet{Thoul_ea:1994}. 
Convection is described using the MLT \citep{BV58}.
We adopt the \citet{Grevesse_Sauval:1998} value of the solar metallicity, $(Z/X)_\odot = 0.023$.

The integration of atmospheric layers in YREC requires that the temperature--optical depth ($T$--$\tau$ hereafter) relation is specified.
While the \citet{KrishnaSwamy:1966} (KS in the following) or Eddington grey approximations give reasonably accurate results for the Sun or more massive stars, respectively, this is not true for the FC models (see section~2.5 of \citealt{Chabrier_Baraffe:1997}, and also Section~\ref{sec:ttau} below).
For this reason, in all the modelling discussed here, we used the $T$-$\tau$ relations taken from the NG model atmospheres of appropriate $(\log g, T_{\rm eff})$.

As a reference, with the input physics described so far, a standard solar model calculated with YREC has an initial helium content of $Y_0=0.278$ and a value of the MLT parameter of $\alpha = 1.82$ ($\alpha = 2.14$) when using the Eddington (KS) $T$--$\tau$ relation. 

\begin{figure}
\begin{center}
\includegraphics[width=0.45\textwidth]{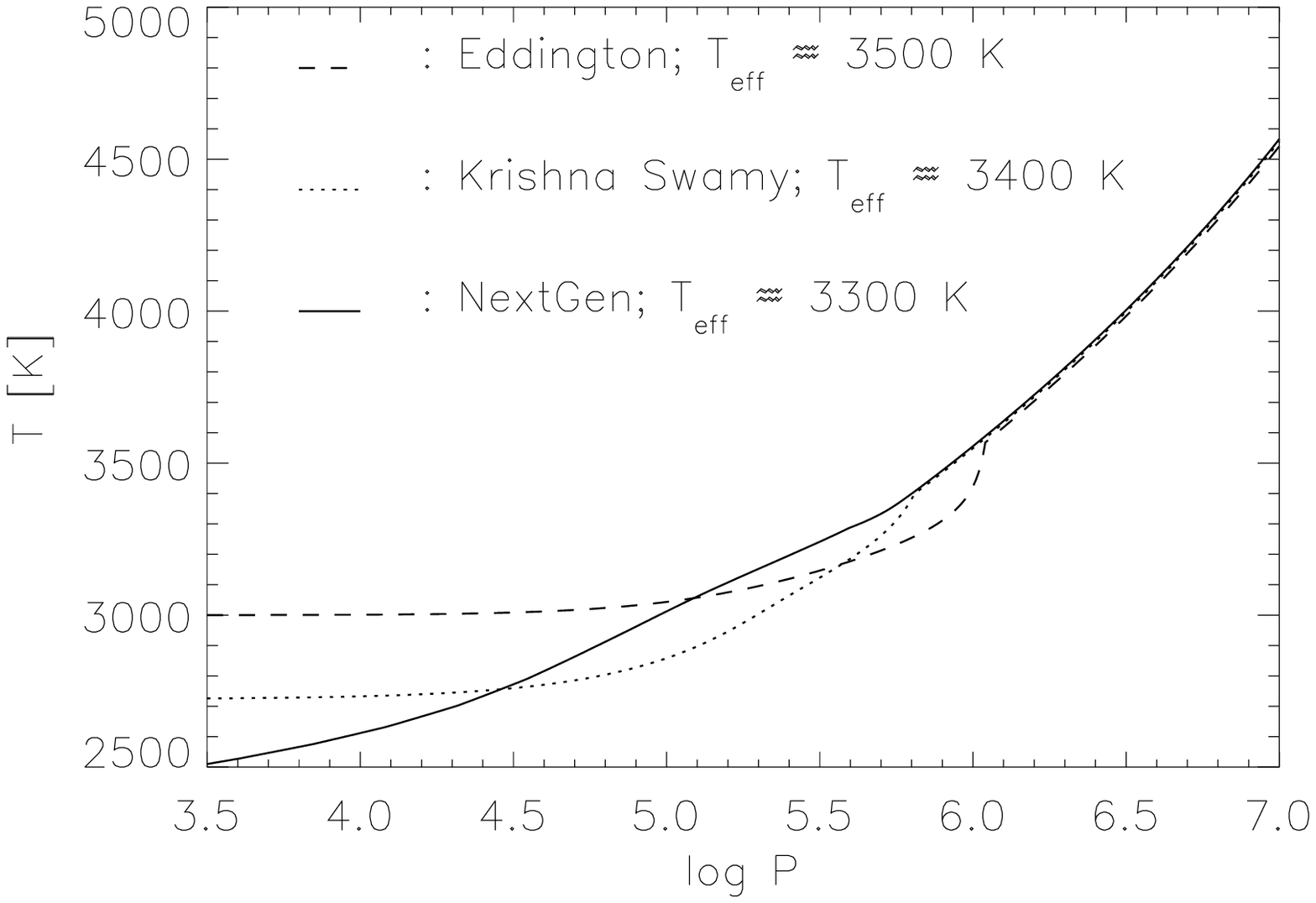}
\includegraphics[width=0.45\textwidth]{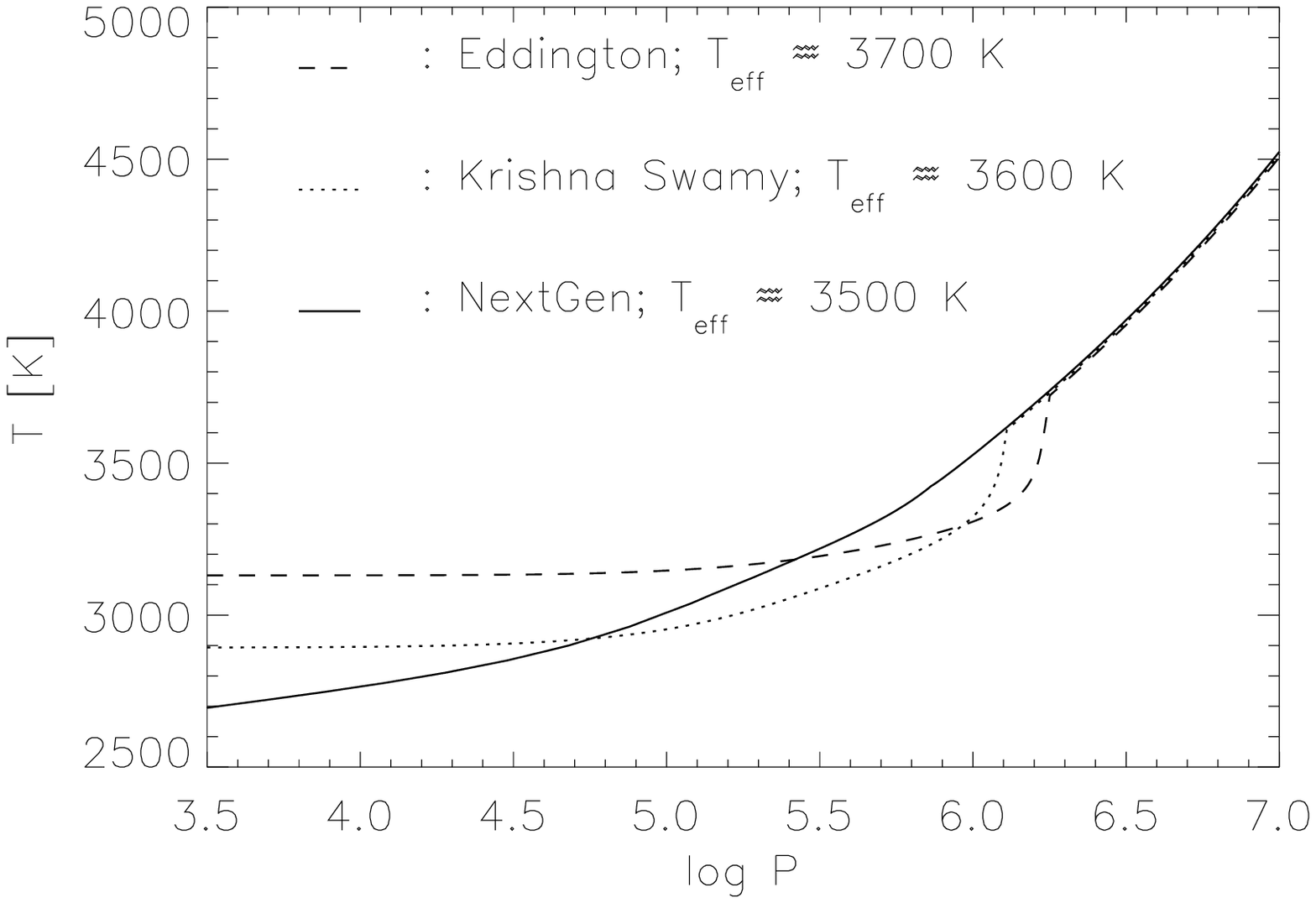}
\caption{Upper panel: $P$-$T$ atmospheric profile for a $0.2~M_\odot$, $4$ Gyr-old star, calculated with $Y=0.28$, $[M/H]=0.0$, and $\alpha=2.0$. Different $T$--$\tau$ relations are used in the atmosphere. Lower panel: the same as in the upper panel, but with $[M/H]=-0.5$.}
\label{fig:ptplane}
\end{center}
\end{figure}

\subsection{The impact of the $T$-$\tau$ relation on FC models}
\label{sec:ttau}

The global parameters of a FC star (i.e. $R$, $L$, $T_{\rm eff}$) are highly sensitive to the treatment of the outermost layers.
In particular, the model atmosphere used to determine the surface boundary conditions plays a crucial role.
As was pointed out by \citet{Chabrier_Baraffe:1997}, a $T$--$\tau$ relation based on a grey atmosphere and/or radiative equilibrium is incorrect for $T_{\rm eff} \lesssim 5000$ K, because the recombination of molecular hydrogen in the envelope reduces the entropy and the adiabatic gradient, thus favouring the penetration of convection in optically thin layers.
Moreover, a grey treatment produces cooler atmospheric profiles below the photosphere, thus leading to overestimated $T_{\rm eff}$ \citep{Chabrier_ea:1996}.

We investigated the impact of these effects on our calculations by comparing models of FC stars constructed using the Eddington, KS and NG $T$--$\tau$ relations. 
Fig. \ref{fig:ptplane} shows the pressure--temperature profile in the outer layers of a star of $0.2~M_\odot$ with solar and moderately metal-poor composition (cf. fig. 5a-b of \citealt{Chabrier_Baraffe:1997}). 
The models calculated with the grey $T$--$\tau$ relations have a discontinuity in the temperature gradient near the photosphere (i.e. around $T = T_{\rm eff}$). 
This is due to the abrupt truncation of convection at the transition from the envelope to the atmosphere, where radiative equilibrium is artificially enforced throughout.
The models using the NG temperature profile, on the other hand, have a much shallower temperature gradient and a smooth envelope-atmosphere transition.
The net effect is that, using Eddington (KS) grey atmosphere models, we obtain effective temperatures which are overestimated of about $200$ K ($100$ K) with respect to the NG models.  

Convection is treated according to the standard MLT in the PHOENIX 1D code. 
Since the models in the NG grid were all constructed with $\alpha=1.0$, we were unable to estimate directly the impact of the value of $\alpha$ in the atmosphere on the global stellar quantities.
Moreover, a value of $\alpha=1.0$ in the atmosphere is larger than the one needed to achieve a satisfactory fit of spectral line profiles (\citealt{VVM_Megessier:1996}, see also section~2 of \citealt{Piau_ea:2011}).
In fact, inconsistent values of $\alpha$ between the atmosphere and the interior of the same stellar model are often required to match satisfactorily all the available observational constraints. 
Numerical simulations of convection also show that assuming the same value of the mixing length throughout all the outer stellar layers might be an oversimplification \citep[see, e.g.][]{Tanner_ea:2011}.
In our modelling, however, this inconsistency does not seem to affect significantly the temperature profile, as is apparent from Fig. \ref{fig:ptplane}.

\section{Results}

\subsection{Modelling the system KOI-126}
\label{sec:k126}

For the system KOI-126, we tried to achieve a self-consistency in the best-fitting models of the three components, as follows.
Given its mass and the preliminary estimate of the age of the system reported by \citet{Carter_ea:2011} of $4\pm1$ Gyr, KOI-126 A is probably in the early post-main sequence phase, on the way to the red giant branch.
Helium and heavy element diffusion and core overshooting were taken into account when modelling this star.
With fixed values of the core overshooting parameter $\alpha_{\rm ov}$ and metal content $[M/H]$, we adjusted the helium content and the MLT parameter in order to produce the model of KOI-126 A with the closest agreement with the observed $T_{\rm eff,A}$ and $R_A$ (we choose to use the $T_{\rm eff}$-$R$ plane to take advantage of the high precision in the radius determination).
The resulting best-fitting model of the primary star gives the most stringent constraint on the age of the system, since the low-mass components undergo very small changes once they are on the main sequence. 
We then tried to fit the radii of KOI-126 B, C for each composition, determining an age range that can be used to check for compatibility with the model of the primary.

In modelling KOI-126 A, which is massive enough to have a convective core, we tried different values of the core overshooting parameter, within the range $\alpha_{\rm ov} = 0.05$--$0.2$. 
This range is typically adopted for early post-main sequence subgiant stars (see, e.g., \citealt{Demarque_ea:2004}, \citealt{Claret:2007}, \citealt{Deheuvels_Michel:2011}).
Interestingly, for this star, we found a very close agreement between the evolutionary tracks calculated with NG ($T_{\rm eff}=5800$ K, $\log g = 4.0$) and Eddington (grey) $T$--$\tau$ relations.

Table \ref{tab:k126a} reports the parameters for the best-fitting models. 
With the values of $\alpha_{\rm ov}$, $[M/H]$, and $Y$ held fixed, the MLT parameter $\alpha$ determines the evolutionary path of the star in the $(T_{\rm eff}, R)$ plane.
The range of $\alpha$ shown corresponds to evolutionary tracks that cross the observational box in the $(T_{\rm eff}, R)$ plane, i.e., it corresponds to the models that passed within $1\,\sigma$ or less from both $R_A$ and $T_{\rm eff,A}$. 
Since the radius determination is much more precise compared with that of the $T_{\rm eff}$, our age estimate is that of the model which has $R\equiv R_A$.
These intervals of $\alpha$ and age do not have the meaning of formal error bars, but are reported here to give a quantitative idea of the sensitivity of the models to the various parameters.

In general, $\alpha_{\rm ov}$ influences the best-fitting value of $\alpha$ and therefore has an indirect impact on the age (lower $\alpha$ values result in younger age estimates).
As for the composition, higher helium content produces younger models, while a higher metallicity requires larger values of $\alpha$ and leads to an increase in the age of the models.

\begin{table}
\caption{Best-fitting parameters for KOI-126 A (ages in Gyr). The meaning of the $\alpha$ and age ranges is explained in the text.}
\begin{center}
\scalebox{0.85}{
\begin{tabular}{ccccccc}
\hline
& \multicolumn{6}{c}{$[M/H]=+0.07$} \\
 & \multicolumn{2}{c}{$\alpha_{\rm ov}=0.05$} & \multicolumn{2}{c}{$\alpha_{\rm ov}=0.10$} & \multicolumn{2}{c}{$\alpha_{\rm ov}=0.20$} \\
\hline
Y & $\alpha$ & age  & $\alpha$ & age  & $\alpha$ & age \\
0.25 & 1.0--1.3 & 3.6--3.8 & 1.1--1.3 & 3.9--4.0 & 1.3--1.4 & 4.0--4.2 \\ 
0.28 & 1.0 & 2.8 & 1.1--1.3 & 3.0--3.1 & 1.1--1.3 & 3.0--3.2 \\
0.30 & $<$ 1.0 & 2.3 & $<$ 1.2 & 2.5 & $<$1.3 & 2.6 \\
\hline
\hline
& \multicolumn{6}{c}{$[M/H]=+0.15$} \\
 & \multicolumn{2}{c}{$\alpha_{\rm ov}=0.05$} & \multicolumn{2}{c}{$\alpha_{\rm ov}=0.10$} & \multicolumn{2}{c}{$\alpha_{\rm ov}=0.20$} \\
\hline
Y & $\alpha$ & age  & $\alpha$ & age  & $\alpha$ & age \\
0.25 & 1.5--2.0 & 4.4--4.7 & 1.2--2.0 & 4.2--4.7 & 1.5--2.0 & 4.7--4.9 \\ 
$0.28$ & $\bf 1.0$--$\bf 1.1$ & $\bf 3.1$--$\bf 3.2$ & $\bf 1.2$ & $\bf 3.4$ & $\bf 1.3$--$\bf 1.4$ & $\bf 3.6$--$\bf 3.7$ \\
$0.30$ & $\bf 1.0$--$\bf 1.2$ & $\bf 2.6$--$\bf 2.8$ & $\bf 1.0$--$\bf 1.2$ & $\bf 2.6$--$\bf 2.8$ & $\bf 1.0$--$\bf 1.5$ & $\bf 2.6$--$\bf 3.2$ \\
\hline
\hline
& \multicolumn{6}{c}{$[M/H]=+0.23$} \\
 & \multicolumn{2}{c}{$\alpha_{\rm ov}=0.05$} & \multicolumn{2}{c}{$\alpha_{\rm ov}=0.10$} & \multicolumn{2}{c}{$\alpha_{\rm ov}=0.20$} \\
\hline
Y & $\alpha$ & age  & $\alpha$ & age  & $\alpha$ & age \\
0.25 & $\bf 1.7$--$\bf 2.0$ & $\bf 5.2$--$\bf 5.3$ & $\bf 1.7$--$\bf 2.0$ & $\bf 5.1$--$\bf 5.3$ & $\bf 1.5$--$\bf 2.0$ & $\bf 5.1$--$\bf 5.4$ \\ 
0.28 & 1.2--1.3 & 4.1--4.2 & 1.2--1.5 & 3.7-3.8 & 1.3--1.5 & 3.9--4.0 \\
0.30 & 1.2 & 3.2 & 1.2--1.3 & 3.2 & 1.4 & 3.4 \\
\hline
\hline
\end{tabular}
}
\end{center}
\label{tab:k126a}
\end{table}

We calculated models of the FC components of KOI-126 with each possible choice of ($Y$, $[M/H]$) shown in Table \ref{tab:k126a}, keeping the MLT parameter fixed with the value $\alpha=2.1$.
We did not make any attempt to account for the $T_{\rm eff}$ constraint, since only a very rough estimate is available for these stars.  
Table \ref{tab:k126bc} shows the age interval compatible with the observed radius for each composition.
Consistent solutions (marked in bold in Tables \ref{tab:k126a} and \ref{tab:k126bc}) exist for the following compositions: $[M/H]=+0.15$, $Y=0.28$ and $Y=0.30$; $[M/H]=0.23$, $Y=0.25$; the ages are $3.4\pm0.3$ Gyr, $2.8\pm0.4$ Gyr and $5.2\pm0.2$ Gyr, respectively. 
In conclusion, using the central determination of $[M/H]$ and a value of $Y=0.28$, close to the solar calibration, we obtain a consistent solution for the system, with an estimated age range of $3.1$--$3.7$ Gyr.

\begin{table}
\caption{Age intervals compatible with the observational constraints on the radius for KOI-126 B and C. All the models were calculated with $\alpha=2.1$.}
\begin{center}
\begin{tabular}{ccccccc}
\hline
 & \multicolumn{2}{c}{$[M/H]=+0.07$} \\
Y & KOI-126 B & KOI-126 C \\
\hline
$0.25$ & $>4.5$ & - \\
$0.28$ & $>3.0$ & $>4.2$ \\
$0.30$ & $2.2$--$4.5$ & $>3.0$ \\
\hline
\hline
 & \multicolumn{2}{c}{$[M/H]=+0.15$} \\
Y & KOI-126 B & KOI-126 C \\
\hline
$0.25$ & $>3.5$ &  $>5.0$ \\
$0.28$ & $\bf 2.0$--$\bf 4.5$ & $\bf >2.7$ \\
$0.30$ & $\bf <3.5$ & $\bf 1.5$--$\bf 5.0$ \\
\hline
\hline
 & \multicolumn{2}{c}{$[M/H]=+0.23$} \\
Y & KOI-126 B & KOI-126 C \\
\hline
$0.25$ & $\bf 2.0$--$\bf 5.5$ & $\bf >2.7$ \\
$0.28$ & $<3.7$ & $1.5$--$5.5$ \\
$0.30$ & $<3.0$ & $<4.0$ \\
\hline
\hline
\end{tabular}
\end{center}
\label{tab:k126bc}
\end{table}

\begin{figure}
\begin{center}
\includegraphics[width=0.45\textwidth]{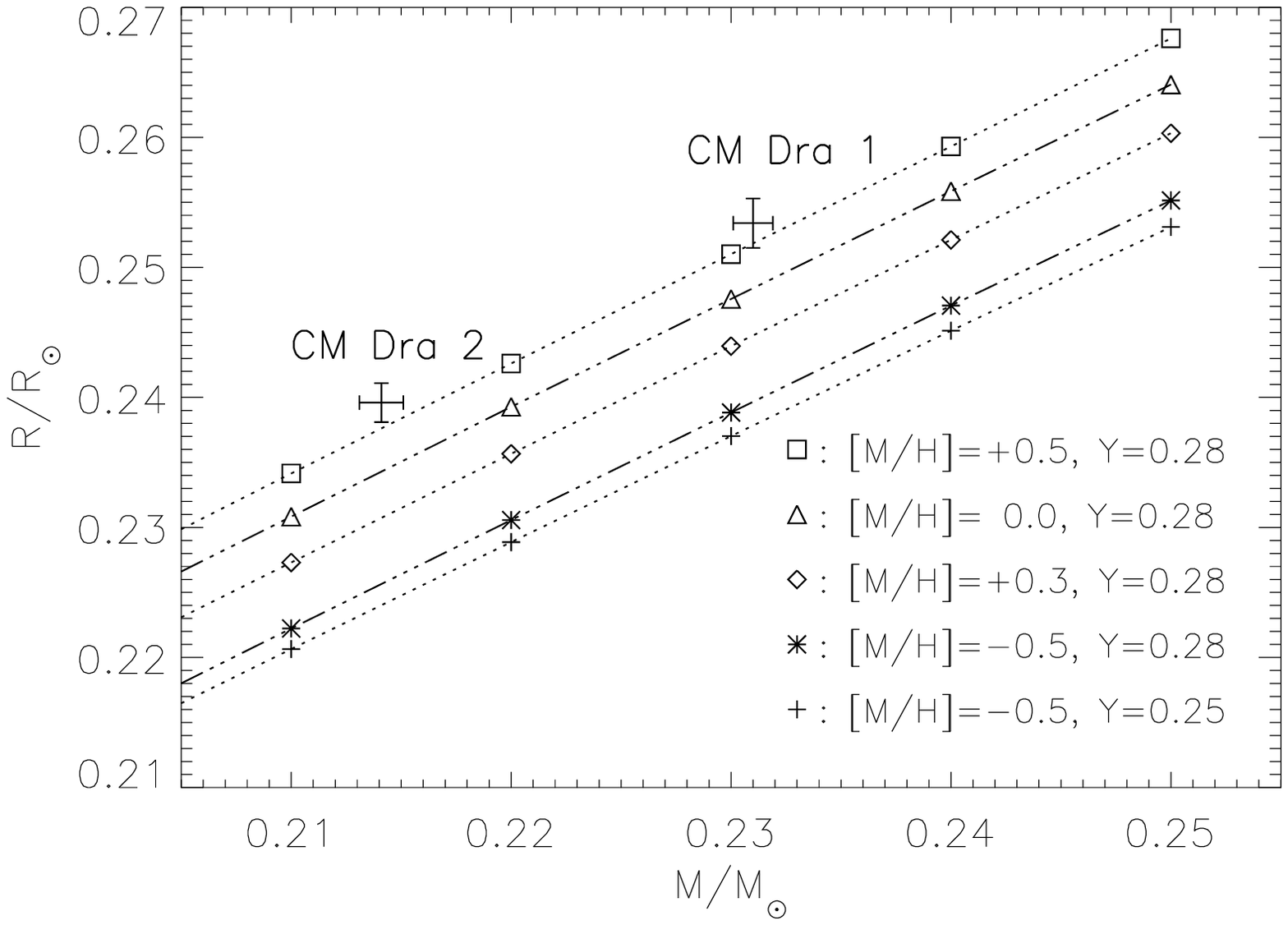}
\includegraphics[width=0.45\textwidth]{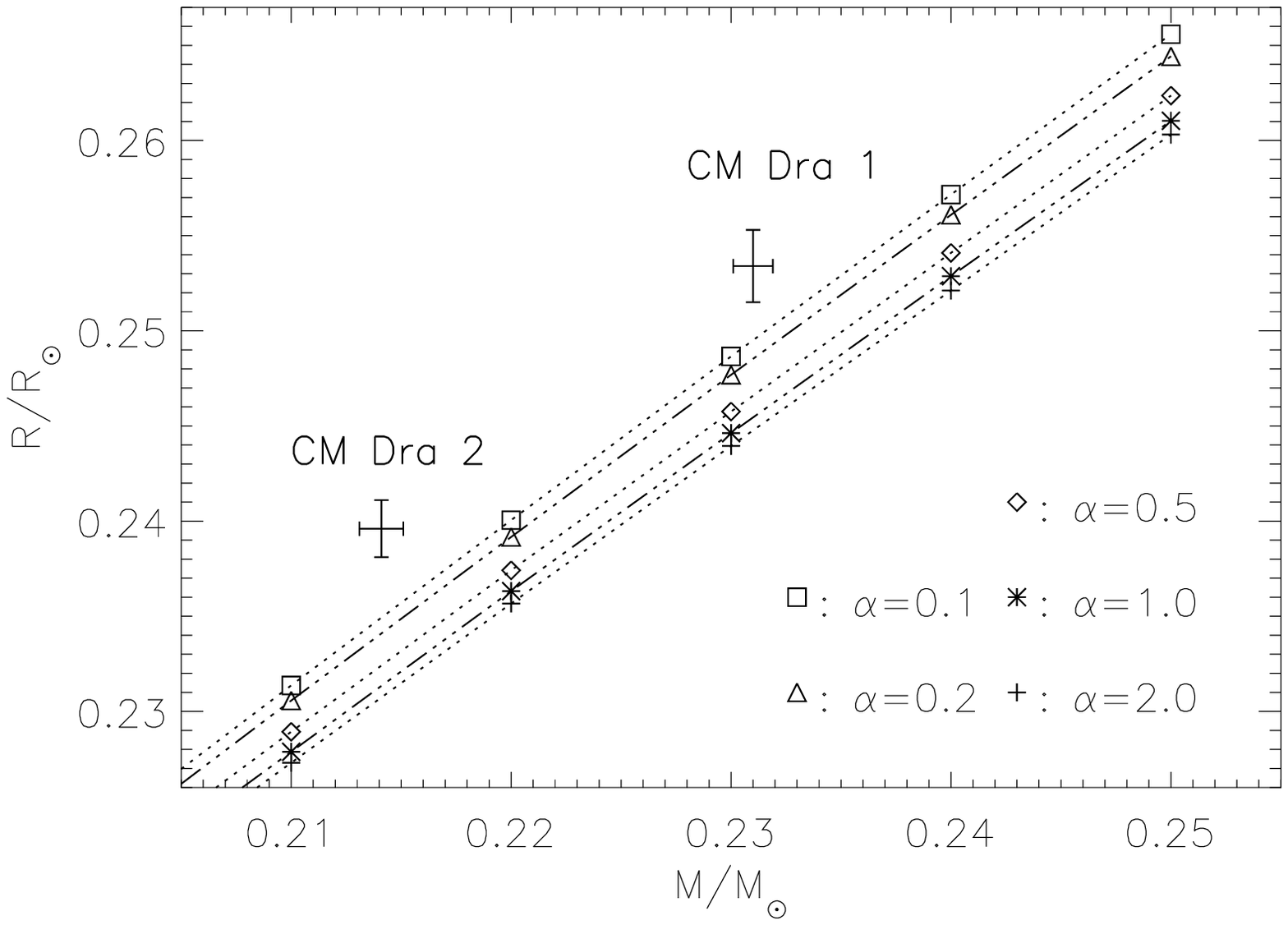}
\caption{Upper panel: comparison of the observational data for CM Dra with theoretical $M$-$R$ relations at $4$ Gyr calculated with different compositions and $\alpha=2.0$. Lower panel: as above, but keeping the composition fixed at solar values (i.e. $Y=0.28$, $Z=Z_\odot$) and using different values of $\alpha$.}
\label{fig:cmdra}
\end{center}
\end{figure}

\subsection{Modelling the system CM Dra}
\label{sec:cmdra}

For CM Dra, we assumed the age quoted in the literature, of about 4 Gyr, and we allowed for a variation of the chemical composition in a very broad range, to take into account the high level of uncertainty in its determination.

As was already clear from Fig.~\ref{fig:mrrel},  it is impossible to reconcile the models of CM Dra with the observed values of the radii when using a metallicity in the range quoted in the literature ($-1.0<[M/H]<-0.6$, see also the upper panel of Fig.~\ref{fig:cmdra}).
In fact, even allowing all the parameters to freely vary within reasonable limits, we always find that our theoretical models underestimate the radii (and overestimate the $T_{\rm eff}$).
The upper panel of Fig.~\ref{fig:cmdra} also shows that, to bring the theoretical models almost within $1\,\sigma$ of agreement with the observations, a rather extreme metal-rich composition must be used ($[M/H]=+0.5$, or $Z = 3~Z_\odot$).
A moderate effect of the helium content is also apparent from the upper panel of Fig.~\ref{fig:cmdra}, with higher values of $Y$ resulting in slightly higher effective temperatures and larger radii.
Such a high metallicity is unlikely for a system of Population II, but it should be emphasized that this membership attribution is based only on the circumstantial evidence provided by the proper motion and that the direct estimate of the metallicity is very controversial \citep[see the discussion in section~5 of][]{Morales_ea:2009}.

The possibility of non-standard effects in the structure of CM Dra components has been invoked in the past to explain the radius anomaly.
Both stars are usually assumed to have synchronised their spin with the orbital period, which makes of them quite fast rotators. 
They are also known to have spotted surfaces (see, e.g. \citealt{Lacy:1977}), which is an indicator of the presence of subsurface magnetic fields.
A fast rotation and/or highly magnetised regime can reduce the efficiency of the convective transport \citep{Gough_Tayler:1966,Tayler:1973}. 
This effect was taken into account by \citet{Chabrier_ea:2007}, as well as a simplified parametrization of the star-spots. 
They found that both effects can be phenomenologically represented by a reduction of the MLT parameter $\alpha$.
We thus tried to construct models of CM Dra with reduced $\alpha$, which suffices to the purpose of investigating such activity-induced effects within the standard theoretical framework, keeping the number of free parameters to a minimum. 
As the lower panel of Fig.~\ref{fig:cmdra} shows, with $Y=0.28$ and solar metallicity, the disagreement can be brought within about $2\, \sigma$ if a value of $\alpha$ as low as $0.2$ is used.
However, even reducing $\alpha$ down to $0.05$, we were unable to construct models with stellar radii compatible with the observations within the $1\, \sigma$ uncertainty.
For $\alpha<0.05$, the MLT treatment implemented in YREC gives rise to numerical instabilities.  
In this sense, tuning $\alpha$ as a free parameter is insufficient, by itself, to reconcile the models of CM Dra 1 and 2 with the observations.
As the moderate sensitivity of FC stars to $\alpha$ requires such a drastic, possibly unphysical reduction, we also tried to model CM Dra using a combination of the composition ($Y$ and $Z$) variation and $\alpha$ reduction effects.
With $[M/H]=+0.3$ (i.e. $Z=2\, Z_\odot$) and $Y=0.28$--$0.30$, for example, we find that the discrepancy is lower than $2\,\sigma$ for $\alpha=1.0$.
Clearly, a better determination of the composition of the system, based on more robust evidences than kinematic arguments, is required to conclusively settle this issue.
Since this information is currently missing, we refrain from attempting a precise modelling of CM Dra, as was done in Section~\ref{sec:k126} for KOI-126. 
We believe, nonetheless, that our results suggest a revision of the metallicity of this system towards at least solar-like (if not higher) values. 

\subsection{Comparison with previous work}
\label{sec:comp}

\begin{figure}
\begin{center}
\includegraphics[width=0.45\textwidth]{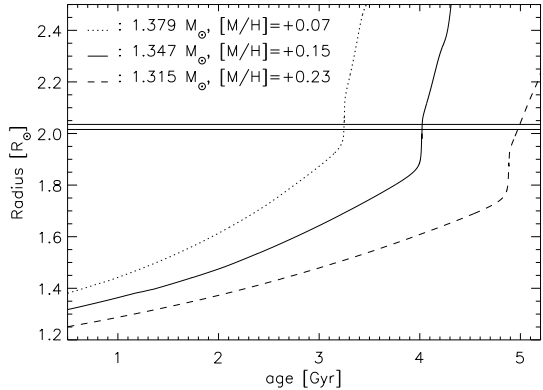}
\caption{Evolutionary tracks for the radius of KOI-126 A compared with the observed range (thin horizontal lines). The resulting ages are $3.2$, $4.0$ and $5.0$ Gyr.}
\label{fig:comp}
\end{center}
\end{figure}

 Dedicated models for the system KOI-126 were constructed, independently of the present work, by \citet{Feiden_ea:2011}, who used the Dartmouth Stellar Evolution Program (DSEP, see \citealt{Chaboyer_ea:2001} and references therein).
They were able to match very well the observational constraints and to reproduce satisfactorily the mass--radius relationship for the FC components, using a metallicity in the quoted range and a solar-calibrated value of the MLT parameter. 
Their estimate for the age of the system is $4.1\pm0.6$ Gyr. 
The codes DSEP and YREC have a distant common origin; they differ mainly in the treatment of some pieces of input physics, namely, the equation of state, the atmospheric boundary conditions, and the presence of turbulent diffusion. 
As was shown in the previous section, both codes are quite successful in modelling the system KOI-126; the differences between the two codes manifest themselves in the estimated age.
To ease the comparison, Fig.~\ref{fig:comp} shows the evolutionary tracks for the radius of KOI-126 A, constructed with YREC using the solar-calibrated MLT parameter and helium content ($Y_0=0.278$, $\alpha=1.82$, Eddington $T$--$\tau$ relation, $\alpha_{\rm ov}=0.2$), and the same choice of mass and metallicity as in  fig.~1 of \citet{Feiden_ea:2011}. 
In this way, we arrive at the age range $4.0\pm 1$ Gyr, with a slightly larger uncertainty, but compatible with the results of \citet{Feiden_ea:2011}.  
Interestingly, our tracks in Fig.~\ref{fig:comp} intersect the observed radius when the star is already out of the main sequence phase, while this is not the case for the models in fig.~1 of \citet{Feiden_ea:2011}.  

\section{Discussion and Conclusions}
\label{sec:disc}

We modelled the systems KOI-126 and CM Dra with the YREC code, using up-to-date input physics.
Both systems have very precise determination of stellar masses and radii.
This allows to test in detail the theoretical models of the FC components, which is the main goal of this work.

As Fig.~\ref{fig:mrrel} shows, for KOI-126 B and C the theoretical mass--radius tracks at $4$ Gyr in the $0.20$--$0.25$ mass range, with a metallicity between solar and $[M/H]=+0.30$ ($Y=0.28$ and $\alpha=2.0$), match quite well the observational data, consistently with the detailed modelling discussed in Section~\ref{sec:k126}.

For the system CM Dra the situation is quite different. 
We find significant discrepancies in radii and $T_{\rm eff}$ when using $\alpha\approx 2$ and a metal-poor, solar or even moderately metal-rich composition (see the upper panel of Fig.~\ref{fig:cmdra}).
Dedicated modelling of CM Dra was attempted in the past. 
Early studies used polytropic models (e.g. \citealt{Pacz84}); \citet{Chabrier_Baraffe:1995}, using a complete stellar modelling approach, found a good agreement between their models (with $[M/H]=-0.5$ and $Y=0.25$) and the observations of \citet{Lacy:1977}. 
The more precise data available today, however, cannot be reconciled with the models of \citet{BCAH98}, the discrepancy being worse for a metal-poor composition with respect to a solar composition \citep{Morales_ea:2009}.
Modelling non-standard input physics, such as stellar activity and/or magnetic fields, has been attempted as well \citep{Chabrier_ea:2007}. 
These effects, however, need not be the only (nor the dominant) explanation. 
The composition, in fact, which is very poorly constrained, can have a significant impact on $R$ and $T_{\rm eff}$ \citep{Burrows_ea:2011}.

In conclusion, our models match satisfactorily the observational data for all the components of KOI-126. 
With $[M/H]=+0.15$ and $Y=0.28$, we obtain an age of about $3.6$ Gyr (based mainly on the modelling of KOI-126 A), which is compatible with the initial estimates of \citet{Carter_ea:2011} and with the results of \citet{Feiden_ea:2011} within the errors.
We are unable, on the other hand, to reconcile our models with the observed radii and $T_{\rm eff}$ of CM Dra if we assume the currently estimated metal-poor composition.
We find that the MLT parameter has a modest effect on the structure of FC stars, which is nevertheless significant with respect to the very precise measurements available for these systems for $\alpha \lesssim 1.0$. 
For example, the relative change in radius for a star of $0.24\, M_\odot$ is up to $2.5$ per cent when $\alpha$ goes from $2.0$ to $0.1$. 
The agreement of CM Dra can therefore be improved using both higher metallicity and a reduced $\alpha$ (e.g. it can be brought within $2\,\sigma$ for $Z= Z_\odot$ and $\alpha = 0.2$ or  $Z= 2\, Z_\odot$ and $\alpha= 1.0$).

\section*{Acknowledgements}
The authors are grateful to S. Sofia for reading the manuscript and providing valuable comments and to B. Chaboyer for private communication on his work in advance of publication. 
FS acknowledges support from the Yale Institute for Biospheric Studies through the YIBS Postdoctoral Fellowship.

\bibliographystyle{mn2e}

\begin{thebibliography}{99}

\bibitem[Allard et al.(1997)]{Allard_ea:1997} Allard, F., Hauschildt, P.~H., Alexander, D.~R., \& Starrfield, S.\ 1997, \araa, 35, 137 

\bibitem[Bahcall \& Pinsonneault(1992)]{Bahcall_Pinsonneault:1992} Bahcall, J.~N., \& Pinsonneault, M.~H.\ 1992, \apjl, 395, L119 

\bibitem[Baraffe et al.(1998)]{BCAH98} Baraffe, I., Chabrier, G., Allard, F., \& Hauschildt, P.~H.\ 1998, \aap, 337, 403 

\bibitem[Berger et al.(2006)]{Berger_ea:2006} Berger, D.~H., et al.\ 2006, \apj, 644, 475 

\bibitem[B{\"o}hm-Vitense(1958)]{BV58} B{\"o}hm-Vitense, E.\ 
1958, Zs. Ap., 46, 108 

\bibitem[Burrows et al.(2011)]{Burrows_ea:2011} Burrows, A., Heng, K., \& Nampaisarn, T.\ 2011, \apj, 736, 47 

\bibitem[Carter et al.(2011)]{Carter_ea:2011} Carter, J.~A., et al.\ 2011, Science, 331, 562 

\bibitem[Chaboyer et al.(2001)]{Chaboyer_ea:2001} Chaboyer, B., Fenton, W.~H., Nelan, J.~E., Patnaude, D.~J., \& Simon, F.~E.\ 2001, \apj, 562, 521

\bibitem[Chabrier \& Baraffe(1995)]{Chabrier_Baraffe:1995} Chabrier, G., \& Baraffe, I.\ 1995, \apjl, 451, L29 

\bibitem[Chabrier \& Baraffe(1997)]{Chabrier_Baraffe:1997} Chabrier, G., \& Baraffe, I.\ 1997, \aap, 327, 1039

\bibitem[Chabrier \& Baraffe(2000)]{Chabrier_Baraffe:2000} Chabrier, G., \& Baraffe, I.\ 2000, \araa, 38, 337 

\bibitem[Chabrier et al.(1996)]{Chabrier_ea:1996} Chabrier, G., Baraffe, I., \& Plez, B.\ 1996, \apjl, 459, L91

\bibitem[Chabrier et al.(2007)]{Chabrier_ea:2007} Chabrier, G., Gallardo, J., \& Baraffe, I.\ 2007, \aap, 472, L17 

\bibitem[Claret(2007)]{Claret:2007} Claret, A.\ 2007, \aap, 475, 1019 

\bibitem[Deheuvels \& Michel(2011)]{Deheuvels_Michel:2011} Deheuvels, S., \& Michel, E.\ 2011, \aap, 535, A91  

\bibitem[Demarque et al.(2008)]{Demarque_ea:2008} Demarque, P., Guenther, D.~B., Li, L.~H., Mazumdar, A., \& Straka, C.~W.\ 2008, \apss, 316, 31 

\bibitem[Demarque et al.(2004)]{Demarque_ea:2004} Demarque, P., Woo, J.-H., Kim, Y.-C., \& Yi, S.~K.\ 2004, \apjs, 155, 667 

\bibitem[Feiden et al.(2011)]{Feiden_ea:2011} Feiden, G.A., Chaboyer, B. \& Dotter, A. \ 2011, \apjl, 740, L25 


\bibitem[Ferguson et al.(2005)]{Ferguson_ea:2005} Ferguson, J.~W., Alexander, D.~R., Allard, F., Barman, T., Bodnarik, J.~G., Hauschildt, P.~H., Heffner-Wong, A., \& Tamanai, A.\ 2005, \apj, 623, 585 

\bibitem[Gough \& Tayler(1966)]{Gough_Tayler:1966} Gough, D.~O., \& Tayler, R.~J.\ 1966, \mnras, 133, 85 

\bibitem[Grevesse \& Sauval(1998)]{Grevesse_Sauval:1998} Grevesse, N., \& Sauval, A.~J.\ 1998, \ssr, 85, 161 

\bibitem[Hansen et al.(2004)]{Kawaler} Hansen, C.~J., Kawaler, S.~D., \& Trimble, V.\ 2004, Stellar interiors: physical principles, structure, and evolution, 2nd ed. New York: Springer-Verlag, 2004

\bibitem[Hauschildt et al.(1999)]{Hauschildt_ea:1999} Hauschildt, P.~H., 
Allard, F., \& Baron, E.\ 1999, \apj, 512, 377 

\bibitem[Iglesias \& Rogers(1996)]{Iglesias_Rogers:1996} Iglesias, C.~A., \& Rogers, F.~J.\ 1996, \apj, 464, 943 

\bibitem[Kippenhahn et al.(1970)]{KMT70} Kippenhahn, R., Meyer-Hofmeister, E., \& Thomas, H.~C.\ 1970, \aap, 5, 155 

\bibitem[Krishna Swamy(1966)]{KrishnaSwamy:1966} Krishna Swamy, K.~S.\ 
1966, \apj, 145, 174 

\bibitem[Lacy(1977)]{Lacy:1977} Lacy, C.~H.\ 1977, \apj, 218, 444 

\bibitem[L{\'o}pez-Morales(2007)]{LopezMorales:2007} L{\'o}pez-Morales, M.\ 2007, \apj, 660, 732 

\bibitem[Metcalfe et al.(1996)]{Metcalfe_ea:1996} Metcalfe, T.~S., Mathieu, R.~D., Latham, D.~W., \& Torres, G.\ 1996, \apj, 456, 356

\bibitem[Morales et al.(2009)]{Morales_ea:2009} Morales, J.~C., et al.\ 2009, \apj, 691, 1400 

\bibitem[Morales et al.(2010)]{Morales_ea:2010} Morales, J.~C., Gallardo, J., Ribas, I., Jordi, C., Baraffe, I., \& Chabrier, G.\ 2010, \apj, 718, 502 

\bibitem[Paczynski \& Sienkiewicz(1984)]{Pacz84} Paczynski, B., \& Sienkiewicz, R.\ 1984, \apj, 286, 332 

\bibitem[Piau et al.(2011)]{Piau_ea:2011} Piau, L., Kervella, P., Dib, S., \& Hauschildt, P.\ 2011, \aap, 526, A100 

\bibitem[Ribas(2006)]{Ribas:2006} Ribas, I.\ 2006, \apss, 304, 89 

\bibitem[Ribas et al.(2000)]{Ribas_ea:2000} Ribas, I., Jordi, C., \& Gim{\'e}nez, {\'A}.\ 2000, \mnras, 318, L55 

\bibitem[Ribas et al.(2008)]{Ribas_ea:2008} Ribas, I., Morales, J.~C., Jordi, C., et al.\ 2008, Mem. Soc. Astron. It., 79, 562 

\bibitem[Rogers \& Nayfonov(2002)]{Rogers_Nayfonov:2002} Rogers, F.~J., \& Nayfonov, A.\ 2002, \apj, 576, 1064 

\bibitem[Tayler(1973)]{Tayler:1973} Tayler, R.~J.\ 1973, \mnras, 165, 39 

\bibitem[Tanner et al.(2011)]{Tanner_ea:2011} Tanner, J., Basu, S., Demarque, P., \& Robinson, F.\ 2011, Journal of Physics Conference Series, 271, 012080 

\bibitem[Thoul et al.(1994)]{Thoul_ea:1994} Thoul, A.~A., Bahcall, J.~N., \& Loeb, A.\ 1994, \apj, 421, 828 

\bibitem[Van Cleve \& Caldwell (2009)]{VanCleve_Caldwell:2009} Van Cleve, J.~E. \& Caldwell, D.~A. 2009, Kepler Instrument Handbook (KSCI-19033) 

\bibitem[van't Veer-Menneret \& Megessier(1996)]{VVM_Megessier:1996} van't Veer-Menneret, C., \& Megessier, C.\ 1996, \aap, 309, 879 

\bibitem[Viti et al.(2002)]{Viti_ea:2002} Viti, S., Jones, H.~R.~A., 
Maxted, P., \& Tennyson, J.\ 2002, \mnras, 329, 290 

\bibitem[Viti et al.(1997)]{Viti_ea:1997} Viti, S., Jones, H.~R.~A., Schweitzer, A., Allard, F., Hauschildt, P.~H., Tennyson, J., Miller, S., \& Longmore, A.~J.\ 1997, \mnras, 291, 780 

\end{thebibliography}

\label{lastpage}

\end{document}